\documentclass[a4paper,final]{appolb}
\usepackage{graphicx}

\usepackage{amsmath}
\usepackage{hyperref}
\usepackage{ulem} % Strikethrough
\usepackage{cancel}
\usepackage{pdfpages}

\graphicspath{{./newfigs/}{./newnewfigs/}}

\begin{document}
% \eqsec  % uncomment this line to get equations numbered by (sec.num)
\title{Algorithmics of Diffraction%
\thanks{Presented at Diffraction and Low-$x$, Reggio Calabria, Italy, August 31st, 2018.}%
% you can use '\\' to break lines
}
\author{Mikael Mieskolainen
\address{Department of Physics, University of Helsinki and Helsinki
Institute of Physics, P.O. Box 64, FI-00014 Helsinki, Finland}
\\
}

\maketitle

\begin{abstract}
We discuss novel ways to probe high energy diffraction, first inclusive diffraction and then central exclusive processes at the LHC. Our new Monte Carlo synthesis and analysis framework, \textsc{Graniitti}, includes differential screening, an expendable set of scattering amplitudes with adaptive Monte Carlo sampling, spin systematics and modern computational technology.
\end{abstract}
%\PACS{PACS numbers come here}
  
\section{Introduction}

High energy diffraction is driven by highly coherent processes at asymptotic energies. One of the main questions is how the unitarity is being composed, related to generalized shadowing and multiparton interactions. In the conventional language, a longstanding problem is the transition between `different' soft and hard QCD Pomerons. Their trajectory intercept $1 + \Delta_P$ controls the cross section $s$-evolution and the trajectory slope $\alpha_P'$ couples with $t$-dependence. In addition, multipomeron exchanges and their interactions beyond elastic, eikonalized scattering, are also relatively unknown. The enigmatic helicity structure of amplitudes is especially interesting in exclusive diffraction. In practise, a single soft Pomeron pole with the effective trajectory parameters, depending on the scheme of its definition and data being fitted, represents at pragmatic level often the most economic parametrization of the bulk of high energy soft scattering data.

Interesting topics beyond the most common include multivariate correlations and fluctuations with stochastic process connections -- Reggeon field theory as a stochastic process and Gribov diffusion \cite{grassberger1978reggeon}, soft exponential and pQCD power law $p_t$-scaling laws, Pomeron via holographic dualities, connections with gravity and high tech scattering amplitude techniques.  Next, we briefly describe some new methods, which can perhaps shed light on some of these topics. Here we emphasize systematic, well-posed but also practical definitions of observables and measurements.

\section{Fiducial vector observables}

The goal for us is to have a self consistent and maximally model independent approach for defining the inclusive diffraction observables. For that, we use the generalized event topology structure, which we have introduced earlier. Combinatorial partial cross sections are now written in a compact tensor product incidence algebra form
\begin{align}
\label{eq:fiducial}
\nonumber
\vec{\sigma} = \frac{1}{2s}
&\sum_f \frac{1}{\text{sym}(f)} \int_{\Omega_f} d\Pi_f \delta^{(4)}\left( p_1+p_2 - \sum_f p_i \right) 
|\mathcal{M}_{2\rightarrow f}|^2
\begin{pmatrix}
 1  &   -1 \\
 0  &   1 \\
\end{pmatrix}^{\otimes \,N} \\
\nonumber
&\begin{pmatrix}
 1  \\
\mathcal{I} \{\Pi_f; \Xi_1 \} \\
\end{pmatrix} \otimes \\
&\hspace{2em}\begin{pmatrix}
 1  \\
\mathcal{I} \{\Pi_f; \Xi_2 \} \\
\end{pmatrix}
\otimes
\cdots
\otimes
\begin{pmatrix}
 1  \\
\mathcal{I} \{\Pi_f; \Xi_N \} \\
\end{pmatrix},
\end{align}
where the acceptance function is $\mathcal{I} : \Pi_f \rightarrow \{0,1\}$, $\Pi_f$ is a set of final state kinematical variables with corresponding Lorentz invariant measure $d\Pi_f$, the symmetry factor is $\text{sym}(f)$  and $\Xi_i$ is the $i$-th fiducial acceptance slice parametrization with $i = 1,\dots,N$. Equation \ref{eq:fiducial} gives us a vector with $2^N$ combinatorial cross section components, essentially also a multivariate polynomial expression or an $N$-point integral correlation functional. A higher rank tensorial structure is obtained when the acceptance function is being treated as a variable function of external kinematics. Also, differential distributions can be embedded in a multiple novel ways within this structure.

This gives us a way of factorizing the model dependent diffractive cross section and Pomeron parameter extractions, and strictly fiducial unfolded vector partial cross sections. In unfolding the detector level quantities back to the particle level, one is typically left with an irreducible residual dependence on the underlying Monte Carlo model driving the detector simulation. The magnitude of this residual depends on how informative the concrete final state observables are. Reconstructed tracks give minimal model dependence, whereas low granularity forward shower counter hits generate larger dependence, as an example. The fiducial definitions should follow closely the active detector acceptance volume in terms of final state pseudorapidity and transverse momentum, ideally.

Thus, by using an $N$-dimensional (multivariate) construction, we have solved the problem of how to make fiducial, future proof generalized measurements of inclusive diffraction. Also, based on fiducial cross sections, we have developed a fast and precise algorithm to fit and extract simultaneously the effective Pomeron intercept and single, double and non-diffractive minimum bias cross sections. Model distributions used in the fit are provided by the chosen Monte Carlo event generator. More extensive MC tunings are also possible.

\begin{figure}[ht]
\centering
\includegraphics[scale=0.47]{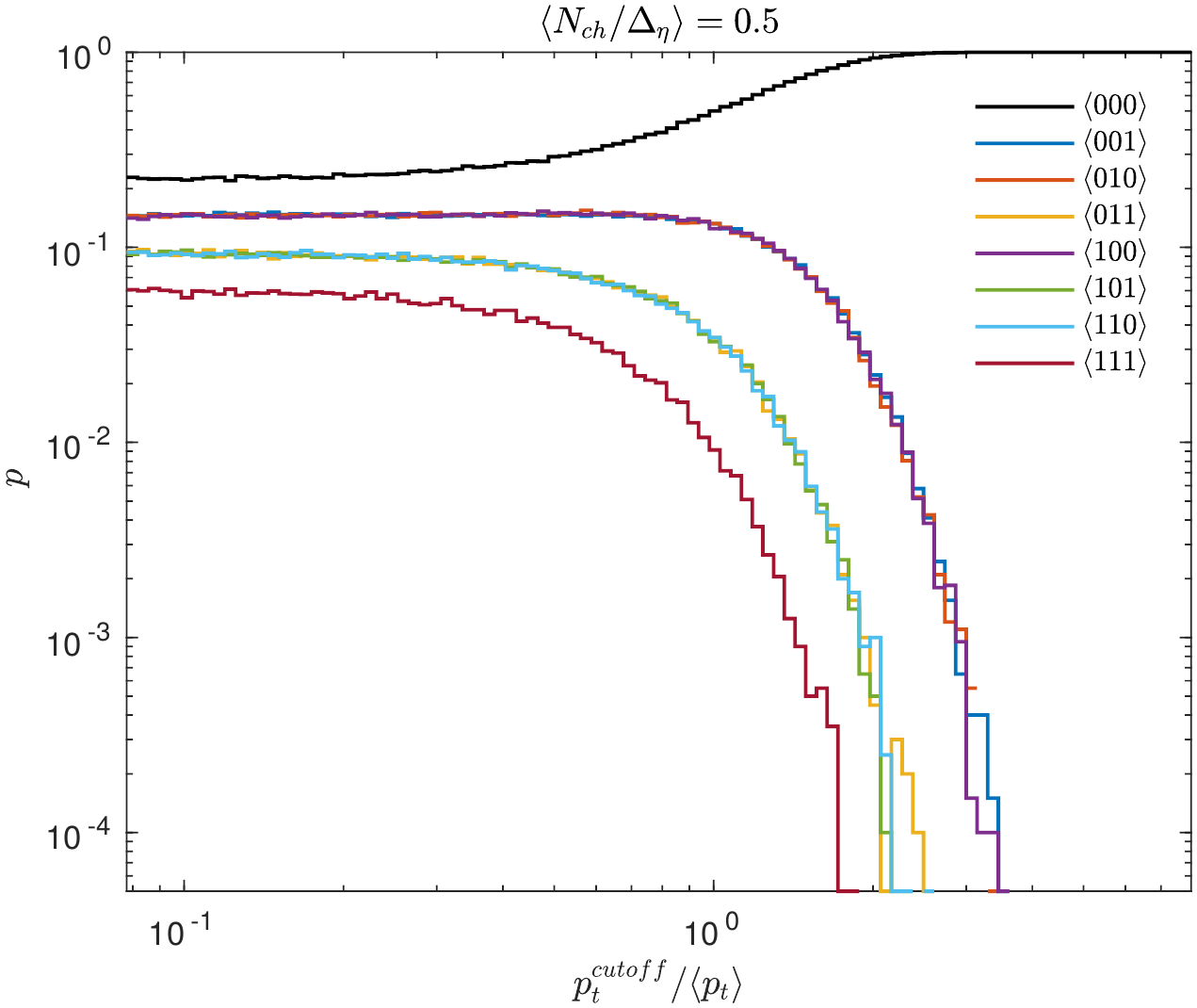}
\includegraphics[scale=0.47]{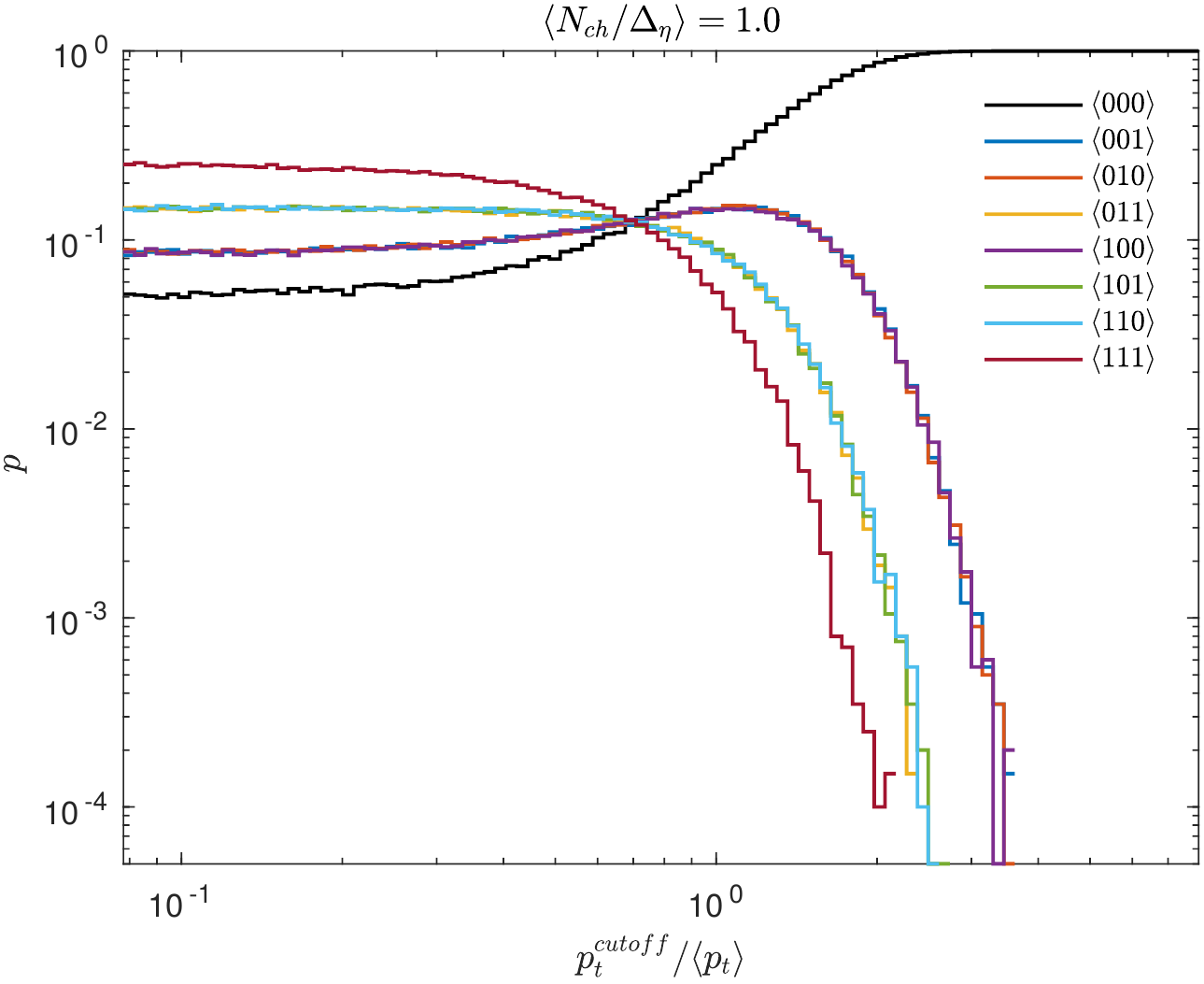}
\includegraphics[scale=0.47]{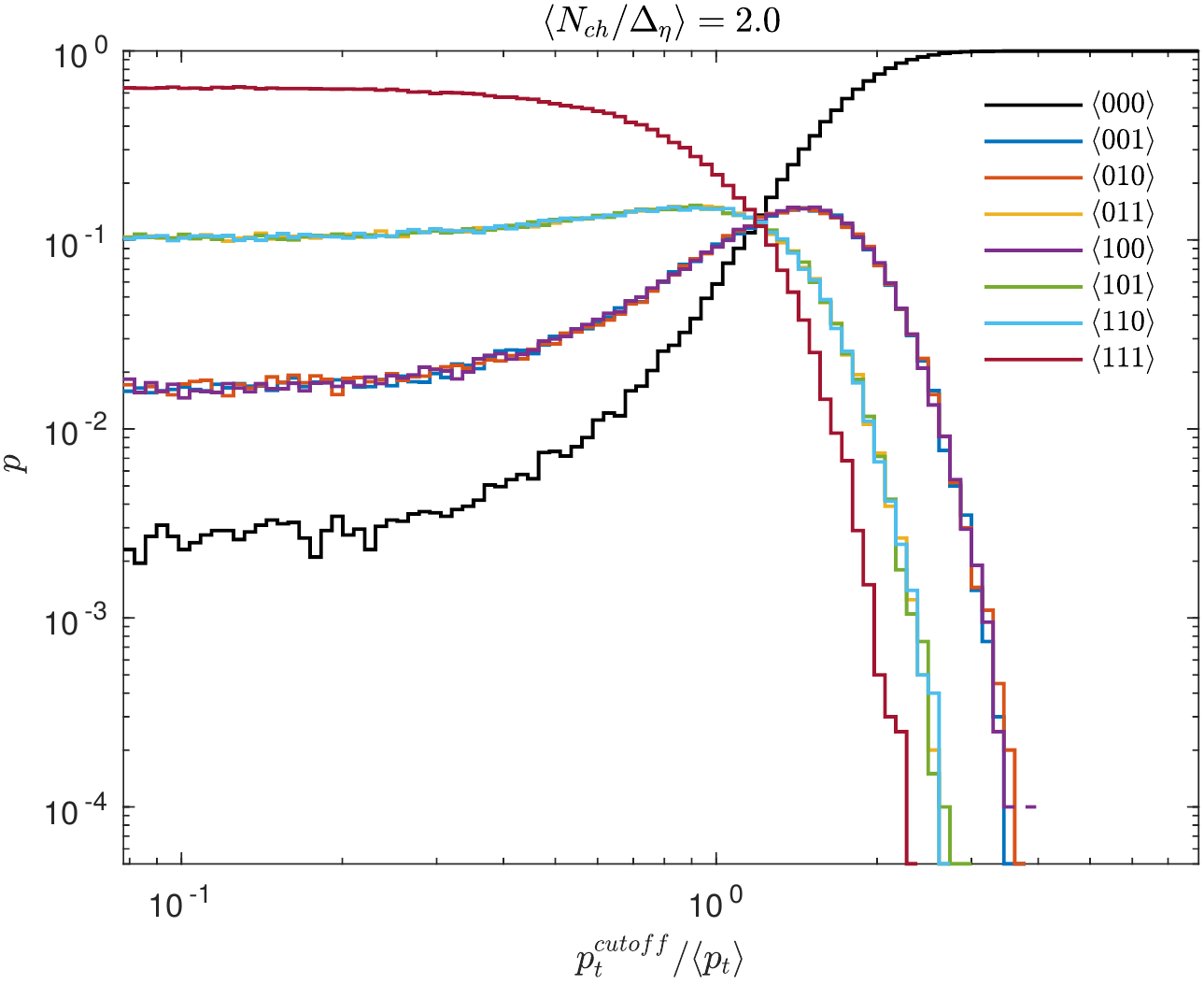}
\includegraphics[scale=0.47]{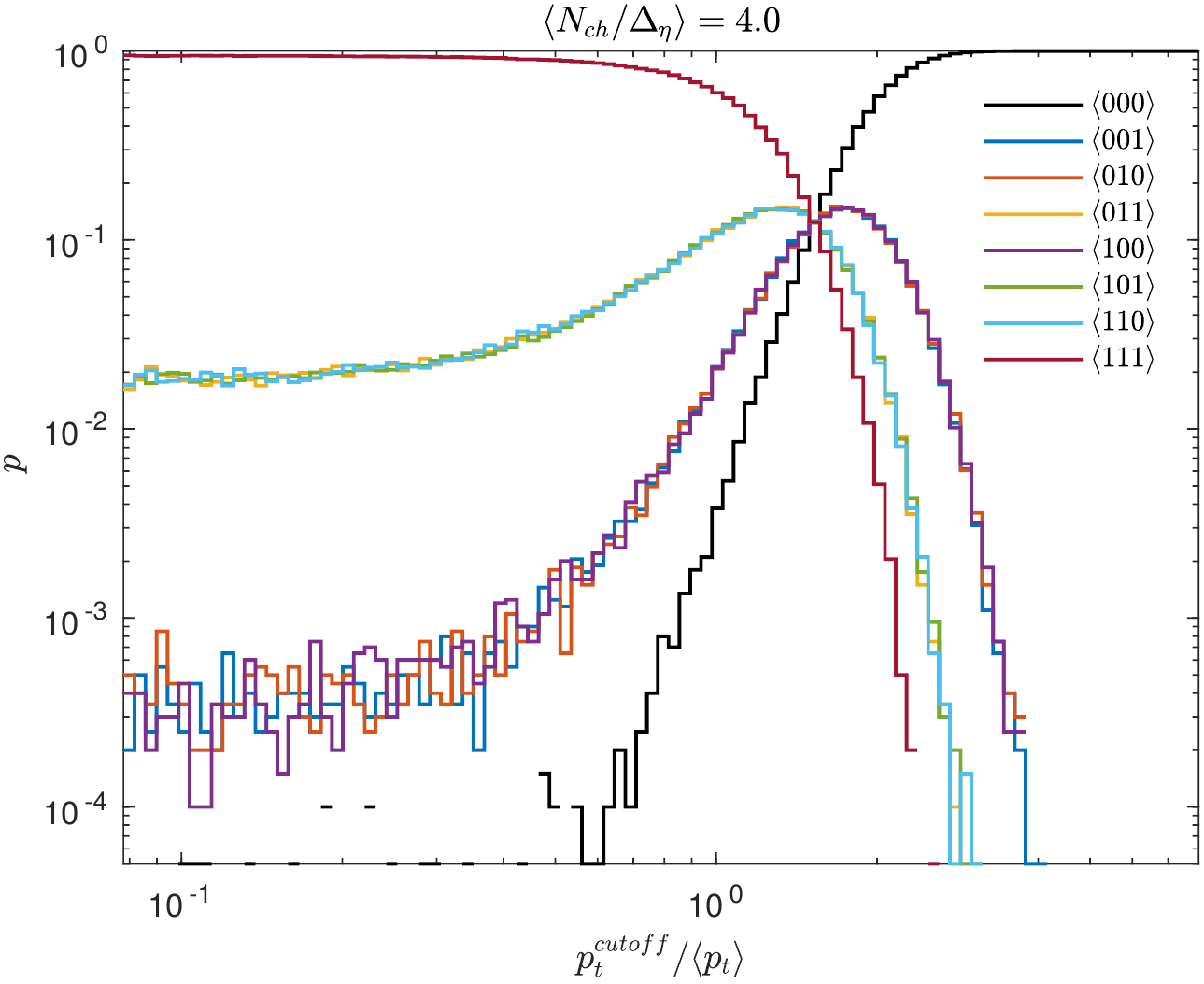}
\caption{Vector valued cross sections with four different particle densities.}
\label{fig: example}
\end{figure}

In Figure \ref{fig: example} we illustrate the partial cross sections, here normalized to probabilities, with a synthetic toy Monte Carlo example in 1+1 dimensions, over rapidity and transverse momentum of the final states. We slice the rapidity dimension into $N = 3$ non-overlapping intervals giving us Bernoulli combinations denoted with $\langle 000 \rangle$, $\langle 001 \rangle$, $\langle 010 \rangle$, $\dots$, $\langle 111 \rangle$. The final state particles are drawn uniformly over rapidity, with fluctuating number of particles per interval following Poisson with mean $\langle N_{ch}/\Delta_\eta \rangle$, with transverse momentum following $ p_t \exp(-b p_t^2)$, which propagates from Gaussian $(p_x,p_y)$-components.

We vary smoothly the $p_t$-cutoff, normalized by $\langle p_t \rangle$, for four different particle densities per discrete rapidity interval $\Delta\eta$, and observe the flow of event topology properties. This demonstration shows that without explicit $p_t$-cutoff descriptions such as experimental characterization, detector simulations and efficiency loss induced rapidity gap data unfoldings, observables relying on rapidity gap structure are not inherently stable.

The vector valued fiducial cross sections absorb the pseudorapidity gap $d\sigma/d\Delta\eta$ distributions of several diffferent type, which can be recovered using a reconstruction algorithm. Expressed in a more mathematical terms, the vector cross sections provide an overcomplete, non-orthogonal basis coefficients to reconstruct other observables. We shall return to this in more detail. Open source code will be made available.

\section{Graniitti: a new Monte Carlo framework}

We have built a new Monte Carlo framework, \textsc{Graniitti}, for simulating and analyzing exclusive diffraction in detail, with maximal flexibility. For completeness, we mention other exclusive generators on the market: \href{https://arxiv.org/abs/1102.2531}{\textsc{FPMC}}, \href{https://arxiv.org/abs/1810.06567}{\textsc{SuperChic}}, \href{https://arxiv.org/abs/1312.4553}{\textsc{Dime}}, \href{https://arxiv.org/abs/1411.6035}{\textsc{GenEx}}, \href{https://arxiv.org/abs/1704.04387}{\textsc{ExDiff}}, \href{https://arxiv.org/abs/1808.06059}{\textsc{CepGen}}, \href{https://arxiv.org/abs/1607.03838}{\textsc{Starlight}}, but also general purpose generators such as \href{https://arxiv.org/abs/1410.3012}{\textsc{Pythia}}, provide diffractive processes. Our approach is perhaps  slightly different than in other exclusive generators; we have approached the generator design more like building a lightweight operating system for the synthesis and analysis dual operation.
\begin{figure}[ht]
\centering
\includegraphics[width=0.7\textwidth]{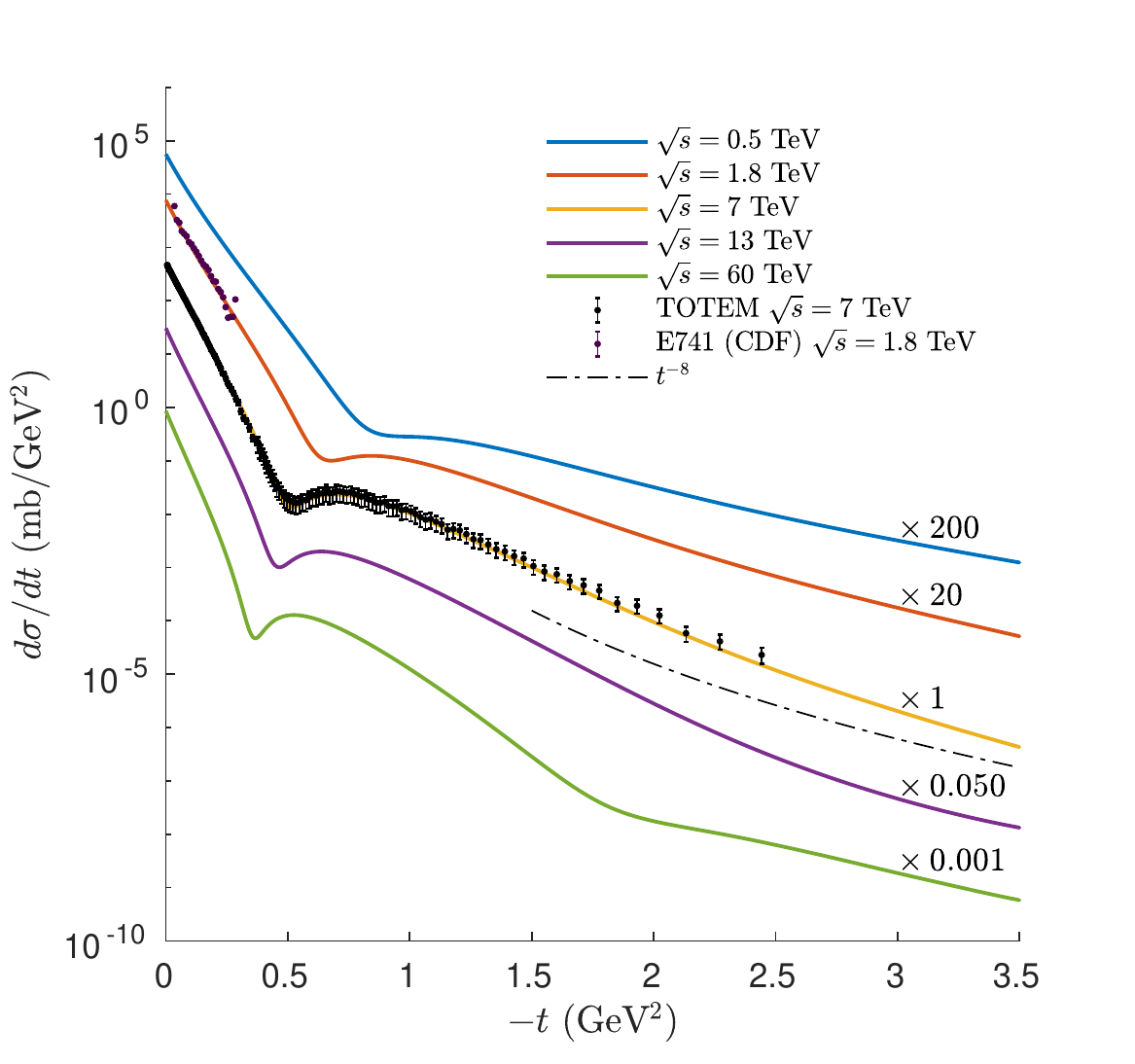}
\caption{Elastic scattering, driving the eikonal differential screening.}
\label{fig: elastic}
\end{figure}

The Monte Carlo sampling machinery uses a fully multithreaded implementation of \textsc{Vegas} Monte Carlo importance sampling. We have been also experimenting with Deep Neural Network based Jacobian transformations for highly efficient Monte Carlo integration and unweighted event generation. After some more work, we expect neural networks to become a part of the event generator toolset. Exact kinematics is provided for all the scattering and decay processes. Arbitrary length decay trees according to phase space are possible, with a fast decay tree syntax interpreter for custom processes. Forward proton low-mass excitation is also equipped with exact kinematics with an adjustable baryonic resonance structure. An interface to \textsc{Pythia} for high mass forward excitation, with suitable excited system parton topologies to attach and span the Lund strings, can be provided.

Elastic $pp$-scattering is implemented via a single channel eikonal model, based on the numerical Fourier-Bessel transform of the single Pomeron exchange amplitude, exponentiation in the conjugate space and the inverse Fourier-Bessel transform. We have studied Odderon amplitudes in the context of new interesting TOTEM results, but have not yet found a minimal description without unnecessary explosion of free parameters.

Soft central production is implemented via double Pomeron continuum amplitudes with two, four and six central final states including symmetrization permutations, similar to \cite{lebiedowicz2010exclusive}. Interfering low-mass resonances are also provided, together with an adjustable spin-parity structure. What is missing is a proper theory to connect all the arbitrary resonance coupling parameters, for transverse and longitudinal degrees of freedom. This is not fully derivable from the first principles at this point, only a very generic spin-parity structure is known \cite{kaidalov2003central}. We do not see that the problem is fully solved neither with fixed vector or tensor current propagator Pomeron models \cite{ewerz2014model}, at least with current data. This topic is of high interest for glueball physics.

For $\gamma\gamma$-processes, we incorporate equivalent photon fluxes with proper photon transverse momentum dependence. Standard Model helicity amplitudes are imported from \href{https://arxiv.org/abs/1106.0522}{\textsc{MadGraph 5}} as \textsc{C++} export, with initial state photons treated on-shell. Via this strategy, we thus enable also arbitrary BSM model amplitude import, using \textsc{UFO} model description language and the automated amplitude generation of \textsc{MadGraph}.

For $gg$-processes, the `Durham model' QCD-framework \cite{harland2014central} is implemented as follows. The main part consists of an event by event numerical 2D-loop integral in $\vec{Q}_t$-space for the screening and fusing gluons with spin-parity projectors together with a pre-calculated skewed transformation of gluon pdf and Sudakov radiation suppression integral factors. Pre-calculations are done automatically, saved, restored and interpolated. Also, we plan to interface this with \textsc{MadGraph} $gg\rightarrow X$ sub-amplitudes. This requires more work than in $\gamma\gamma$-case, due to the more involving QCD color structure implemented in color factorized amplitudes of \textsc{MadGraph}. Essentially, what we need to have is the color singlet projected central state.

Differential loop screening, similar to \textsc{SuperChic} \cite{harland2014central}, is a next-to-leading order feature making impact on cross section normalization (absorption) and especially transverse plane observables, but also other. In the loop screening amplitude we use exactly the same single channel eikonalized pomeron as we use for the elastic scattering. Direct extension would be a multichannel (Good-Walker) eikonals with more free parameters, though. Eikonal screening requires a numerical 2D-loop integral in $\vec{k}_t$-space, event by event. We obtain an integrated screening absorption factor $\langle S^2 \rangle$ at the LHC for $gg \rightarrow gg$ between $0.02-0.03$, for low-mass Pomeron-Pomeron between $0.1-0.2$, for photoproduction of $\rho^0$ between $0.6-0.7$ and in $\gamma \gamma \rightarrow \ell \bar{\ell}$ between $0.85-0.95$, within the phase space accessible in the LHC experiments. Values here are subject to certain changes with more extensive fits of elastic data, but they demonstrate the large dynamic range. We have also automated and transparent fit code for the eikonal model parameters.

We provide arbitrary helicity amplitudes of Jacob-Wick style for low-mass resonances decaying to pseudoscalar pairs, parametrized via von Neumann spin polarization density matrices. For the analysis we have implemented a complete spherical harmonics expansion differentially in the central system kinematics (mass, transverse momentum) in typical Lorentz frames. The code calculates the spherical moment mixing matrices which can be heavily non-diagonal by limited angular acceptance of detectors. An inversion machinery for these is provided via an extended Maximum Likelihood fit and an explicit reqularized algebraic inversion. Care must be taken with angular flat phase space definitions, which are not necessarily strictly fiducial always thus result in extrapolation, a fact to be remembered.

For operation, \textsc{Graniitti} requires standard \textsc{make} tools, a \textsc{C++1x} compatible compiler, \textsc{HepMC3} and \textsc{LHAPDF6} libraries (installer provided). \textsc{Root} libraries are necessary for the fit and analysis tools (optional). The code will be available under the MIT license at \href{https://github.com/mieskolainen}{github.com/mieskolainen}.

\vspace{-0.58em}
\section*{Acknowledgements}

\noindent The organizers of Diffraction and Low-$x$ 2018 are thanked for an excellent conference, and Risto Orava for discussions.

\vspace{-1.25em}
% References
  \let\oldthebibliography=\thebibliography
  \let\endoldthebibliography=\endthebibliography
  \renewenvironment{thebibliography}[1]{%
    \begin{oldthebibliography}{#1}%
      \setlength{\parskip}{0ex}%
      \setlength{\itemsep}{0ex}%
  }%
  {%
    \end{oldthebibliography}%
  }

\nocite{*}
\bibliographystyle{polonica}%
\bibliography{sample}%

\end{document}